\documentclass[usenatbib]{mn2e}
\usepackage{natbibmnfix, graphicx, times, amsmath, epsfig, amssymb}

\newcommand\lsim{\mathrel{\rlap{\lower4pt\hbox{\hskip1pt$\sim$}}
        \raise1pt\hbox{$<$}}}
\newcommand\gsim{\mathrel{\rlap{\lower4pt\hbox{\hskip1pt$\sim$}}
        \raise1pt\hbox{$>$}}}
\def\myputfigure#1#2#3#4#5%
{\vskip#5pt\makebox[0pt]{\hskip#2in
\includegraphics[width=#3\textwidth]{#1}}\vskip#4pt\hfill}

\newenvironment{packed_enum}{
\begin{enumerate}
  \setlength{\itemsep}{1pt}
  \setlength{\parskip}{0pt}
  \setlength{\parsep}{0pt}
}{\end{enumerate}}

\pdfoutput=1

\begin{document}

\title[Depletion of gas in galaxies from reionization]{The depletion of gas in high-redshift dwarf galaxies from an inhomogeneous reionization}
\author[Sobacchi \& Mesinger]{Emanuele Sobacchi$^1$\thanks{email: emanuele.sobacchi@sns.it}, Andrei Mesinger$^1$  \\
$^1$Scuola Normale Superiore, Piazza dei Cavalieri 7, 56126 Pisa, Italy\\
}

\voffset-.6in

\maketitle

\begin{abstract}
The reionization of the intergalactic medium (IGM) was likely inhomogeneous and extended.  By heating the IGM and photo-evaporating gas from the outskirts of galaxies, this process can have a dramatic impact on the growth of structures.
Using a suite of spherically-symmetric collapse simulations spanning a large parameter space, we study the impact of an ionizing ultraviolet background (UVB) on the condensation of baryons onto dark matter halos.  We present an expression for the halo baryon fraction, $f_{\rm b}$, which is an explicit function of: (\textit{i}) halo mass; (\textit{ii}) UVB intensity; (\textit{iii}) redshift; (\textit{iv}) redshift at which the halo was exposed to a UVB.
We also present a corresponding expression for the characteristic or critical mass, $M_{\rm crit}$, defined as the halo mass which retains half of its baryons compared to the global value.  Since our results are general and physically-motivated, they can be broadly applied to inhomogeneous reionization models.
\end{abstract}


\section{Introduction}
\label{sec:intro}

As the first galaxies formed, their ionizing UV radiation carved-out HII regions from the neutral intergalactic medium (IGM).  As structures continued to form, these HII regions grew and overlapped, eventually permeating the entire Universe.  This global phase change, known as reionization, is expected to be fairly extended and inhomogeneous (e.g. \citealt{FHZ04, ZMQT11}).
The resulting ionizing ultraviolet background (UVB) heated the IGM to temperatures of $\sim 10^{4}\text{ K}$, photo-evaporating gas out of shallow potential wells and affecting its cooling properties (e.g. \citealt{SGB94,MR94, HG97}).  Therefore the gas reservoir available for star formation is decreased for low-mass galaxies in the ionized IGM (e.g. \citealt{CR86, Rees86, Efstathiou92}).

This UVB feedback mechanism can have several important effects:
(\textit{i}) by suppressing star-formation inside cosmic HII regions, it results in a more uniform reionization morphology (e.g. \citealt{QLZD07});
(\textit{ii}) it can delay the completion of reionization, helping to alleviate potential tension between a reionization with a mid-point at $z\sim10$ \citep{WMAP11}, and an end-point at $z\sim$6--7 (e.g. \citealt{Mortlock11, Pentericci11, DMW11, BHWH11, SMH12, Ono12});
\footnote{Note that extended reionization scenarios could also result from feedback from early sources of X-rays \citep{RO04, MFS12},
 or from reionization ``stalling'' when the HII regions grow to surpass the attenuation length of ionizing UV photons \citep{FM09, CMMF11, AA12}.}
(\textit{iii}) it can suppress the baryon content of local dwarf galaxies, explaining the apparent dearth of observed satellite galaxies in the Milky Way (e.g. \citealt{KKVP99, MGGL99, BKW00, Somerville02}) and dwarf galaxies in the field (as inferred from the HI ALFALFA survey; \citealt{PMGH11, FANS12}).

Since semi-analytic calculations (e.g. \citealt{SGB94, GH98}) were used to motivate a suppression of star formation in galaxies with $M\lesssim 10^{9}M_{\odot}$, several attempts have been made to further quantify this effect, both using spherically symmetric simulations \citep{TW95, KTUS00, DHRW03} and three-dimensional cosmological hydrodynamic simulations (e.g. \citealt{QKE96, WHK97, NS97, Gnedin00, HYGS06, OGT08}).
\citet{Gnedin00} (see also \citealt{NBM09}) used numerical simulations to argue that the resulting decrease of baryonic content of halos was well predicted by the linear theory filtering mass \citep{GH98}. However, subsequent simulations showed that this prescription should to be revised in the low-redshift, post-reionization regime \citep{HYGS06, OGT08}, taking into account the cooling properties of gas at the outskirts of galaxies.

Unfortunately, quantifying UVB feedback is challenging. Semi-analytic approaches do not include all of the relevant physics (e.g. non-linear growth of perturbations, cooling processes of the gas).  On the other hand, cosmological simulations often focus on a particular reionization history, fixing the UVB evolution and/or the reionization redshift.
Our poor understanding of the details of reionization motivates a flexible approach that is not dependent on a particular model, and can therefore be broadly applied.

As an intermediate approach, in this paper we run suites of fast, 1D cosmological collapse simulations, exploring a wide parameter space motivated by the inhomogeneity of reionization. Fitting these results, we present an expression for the halo baryon fraction, $f_{\rm b}$, as well as the critical mass $M_{\rm crit}$, defined as the halo mass which retains half of its baryons compared to the global value.

In this work, we focus on halos with masses of $10^8 \lsim M/M_\odot \lsim 10^{10}$, which are relevant in the advanced stages of reionization (e.g. \citealt{HRL97}).  These halos are massive enough to host gas collapsing via the atomic cooling channel, and yet small enough to be susceptible to UVB feedback.
  Furthermore, we do not attempt to model baryonic suppression through internal feedback processes, such as internal radiation or winds (e.g. \citealt{SH03,HQM11}). These additional processes require parameter studies of detailed interstellar medium physics of high-$z$ galaxies; isolating their contribution is best done through a separate line of investigation.

The paper is organized as follows.  In \S \ref{sec:simulation_collapse} we describe the collapse simulations we use. In \S \ref{sec:results} we use results from these simulations to fit formulae for $M_{\rm crit}$ and $f_{\rm b}$,  also highlighting the limitations of our approach.  Finally, in \S \ref{sec:concl} we present the conclusions of the work. Throughout we assume a flat $\Lambda\text{CDM}$ cosmology with parameters ($\Omega_{\rm m}$, $\Omega_{\Lambda}$, $\Omega_{\rm b}$, $h$, $\sigma_{\rm 8}$, $n$) = (0.27, 0.73, 0.046, 0.7, 0.82, 0.96), consistent with WMAP results \citep{WMAP11}.

\section{Hydrodynamical Collapse Simulations}
\label{sec:simulation_collapse}

\begin{figure}
\vspace{+0\baselineskip}
{
\includegraphics[width=0.45\textwidth]{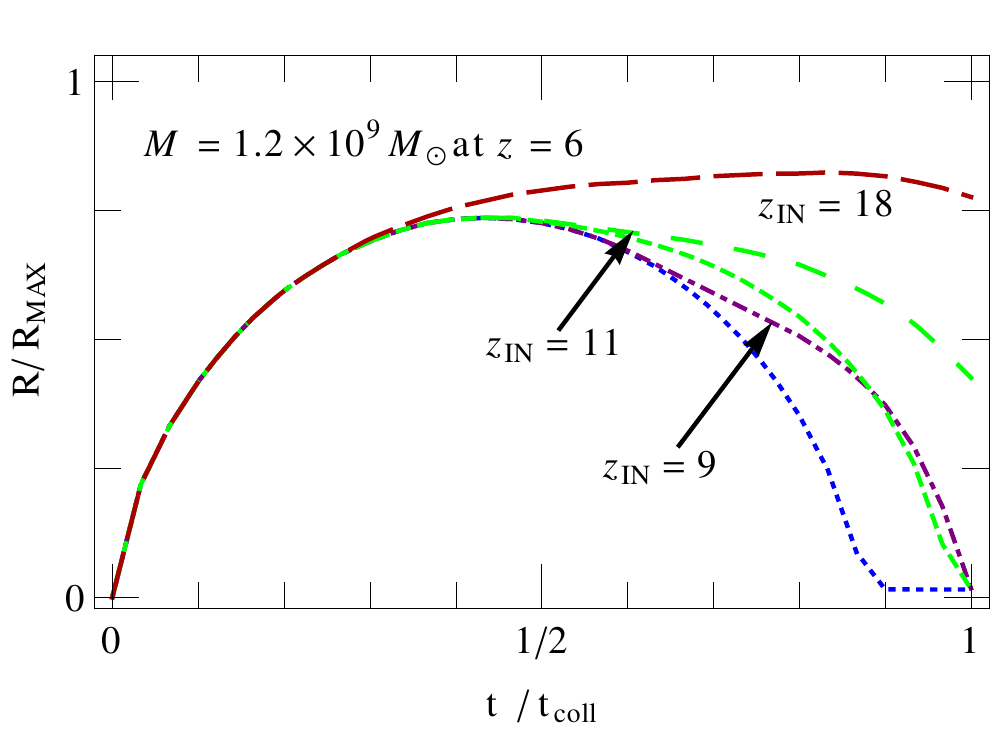}
}
\caption{Evolution of the gas shell enclosing half the baryonic mass of a $1.2\times 10^{9}M_{\odot}$ halo collapsing at $z=6$, for several of our runs. 
The radius is expressed in units of $R_{\rm MAX}$: the maximum radius reached by the {\it outermost} shell (enclosing the total halo mass at $z=6$) in the run without the UVB.  Similarly, the time is expressed in units of $t_{\rm coll}$, the cosmic time at the collapse redshift ($z=6$) of the outermost shell.
 \textit{Dotted}: no UVB. \textit{Dot-dashed}: the UVB (with $J_{\rm 21}=1$) is turned on at $z_{\rm IN}=9$. \textit{Short-dashed}: the UVB (with $J_{\rm 21}=0.01$) is turned on at $z_{\rm IN}=11$. \textit{Dashed}: the UVB (with $J_{\rm 21}=1$) is turned on at $z_{\rm IN}=11$. \textit{Long-dashed}: the UVB (with $J_{\rm 21}=1$) is turned on at $z_{\rm IN}=18$.
\label{fig:shell_evo}
}
\vspace{-1\baselineskip}
\end{figure}

We use a 1D simulation to study collapsing, cosmological perturbations. The code evolves a mixture of dark matter (DM) and baryon fluids by moving concentric spherical shells of fixed mass in the radial direction.  It includes cooling through excitation, ionization, recombination and free-free emission and by Compton scattering with CMB photons (relevant at $z\gtrsim 7$; \citealt{DHRW03}).  For more details on the code please see \citet{TW95}.

As in previous works (e.g. \citealt{TW95, DHRW03, MD08}) we start with typical profile around a 2-$\sigma$ peak\footnote{We find that our results are not very sensitive to the choice of the peak height.  In particular, switching to a 1.5 (2.5) sigma peak  lowers (raises) by $\sim15$\% the derived value of $M_{\rm crit}$.} in a Gaussian random field \citep{BBKS86}, sampled by 6000 (1000) shells for the DM (baryons).
To set-up the runs, we fix the total halo mass $M$ collapsing at redshift $z$. Note that due to the hierarchical nature of structure formation, the halo mass increases with time, so $M$ at $z$ corresponds to a smaller $M_1 < M$ at $z_1 > z$. \footnote{To facilitate comparisons below, we define the halo mass $M$ independently of the UVB. Since the accretion of DM (making up most of the mass) on halo scales is not sensitive to the UVB, this convention effectively translates to $M= (\Omega_{\rm m}/\Omega_{\rm DM}) M_{\rm DM}$.}
We include an isotropic UVB instantaneously turned on at redshift $z_{\rm IN}$ and parametrized as
\begin{equation}
J\left(\nu\right)=J_{\rm 21}\left(\nu/\nu_{\rm H}\right)^{-\alpha}\times 10^{-21}\text{ erg s$^{-1}$ Hz$^{-1}$ cm$^{-2}$ sr$^{-1}$}
\end{equation}
where $\nu_{\rm H}$ is the Lyman limit frequency and $\alpha=5$ corresponds to a stellar-driven UV spectrum (e.g. \citealt{TW96}).
We consider a range of $z_{\rm IN}$= 9--16, as well as three different intensities: $J_{\rm 21}=0.01$, $J_{\rm 21}=0.1$ and $J_{\rm 21}=1$, spanning the expected range of interest \footnote{Measurements of the Ly$\alpha$ forest imply $J_{21}\sim1$--0.1 at $z\sim$2--6 (e.g. \citealt{BH07, CBHB11, HM12}).  One might expect this decreasing trend of the UVB to continue to higher redshifts, since there are fewer collapsed structures and the mean free path of ionizing photons is decreasing with redshift (e.g. \citealt{QOF11, CHS11}).
 However the drop in the number of galaxies could be compensated by an increase in their ionizing efficiency, due to e.g. a higher escape fraction or a top-heavy stellar initial mass function (e.g. \citealt{Schaerer02, WC09, YCN11, FL12}).
  Since the evolution of the UVB intensity is very uncertain at $z>6$, we explore a large range of intensities.}.
This parameter space is motivated by paradigm of inhomogeneous reionization: as reionization proceeds, ionization fronts pass through different regions of the IGM at different redshifts, $z_{\rm IN}$, until they finally overlap at $z\gsim$5--6 (e.g. \citealt{MMF11}).

The effect of the ionizing background on the evolution of a gas shell is shown in Fig. \ref{fig:shell_evo}.
This shell {\it encloses half of the baryonic mass} of a $M=1.2\times 10^{9}M_{\odot}$ halo collapsing at $z=6$; this total halo mass corresponds to, e.g., $M=1.2\times 10^{8}M_{\odot}$ at $t=3t_{\rm coll}/4$ ($z\simeq 7.4$) and $M=2\times 10^{7}M_{\odot}$ at $t=t_{\rm coll}/2$ ($z\simeq 10$). The dotted curve shows the evolution of the shell with no UVB. As expected in the (pressureless) spherical collapse scenario (e.g. \citealt{GG72}), the maximum radius of this shell is reached at half of its collapse time.

The dot-dashed curve shows the evolution of the shell when a $J_{\rm 21}=1$ UVB is turned on at $z_{\rm IN}=9$. The gas is heated by the UVB and the resulting pressure support delays the collapse.  Interestingly this shell (enclosing half of the baryonic mass) collapses at $t \sim t_{\rm coll}$, indicating that $M=1.2\times 10^{9}M_{\odot}$ corresponds to the so-called characteristic or critical mass (see below) for this parameter combination: $z$, $z_{\rm IN}$ and $J_{\rm 21}$.
Similar behavior is shown with the short-dashed curve, corresponding to the run when the background is turned on earlier, $z_{\rm IN}=11$, but with a reduced intensity, $J_{\rm 21}=0.01$: the evolution is delayed since earlier times but the pressure support is lower, therefore this shell also collapses at $t \sim t_{\rm coll}$. These two points in our parameter space, $(M, z, z_{\rm IN}, J_{21})$ = $(1.2\times 10^9 M_\odot, 6, 9, 1)$ and $(1.2\times 10^9 M_\odot, 6, 11, 0.01)$, both result in a baryon content which is $\sim1/2$ of the value expected without a UVB, i.e. $f_b\sim1/2$.

The medium-dashed curve shows the evolution of the shell when $J_{\rm 21}=1$ and $z_{\rm IN}=11$: the collapse is more significantly delayed and the corresponding halo baryon fraction is $f_{\rm b}<1/2$. This is also true for the long-dashed curve, corresponding to the same intensity $J_{\rm 21}=1$ but a higher $z_{\rm IN}=18$. In this case the shell expands slightly after exposure to the UVB and is pressure-supported, with the radius evolving very slowly.  However, as the enclosed potential well deepens from further DM accretion, the shell eventually starts to collapse again at $t \sim t_{\rm coll}$.

The evolution is simpler in smaller halos with $M\sim 10^{8} M_{\odot}$ at $z=6$ (not shown in the figure) since the gas can be more easily photo-evaporated out of the shallower potential wells.  Shells enclosing even half of the mass ``peal-off'' at $z=z_{\rm IN}$, effectively escaping back into the IGM.

\section{Results}

\label{sec:results}

\begin{figure*}
\vspace{+0\baselineskip}
{
\includegraphics[width=0.33\textwidth]{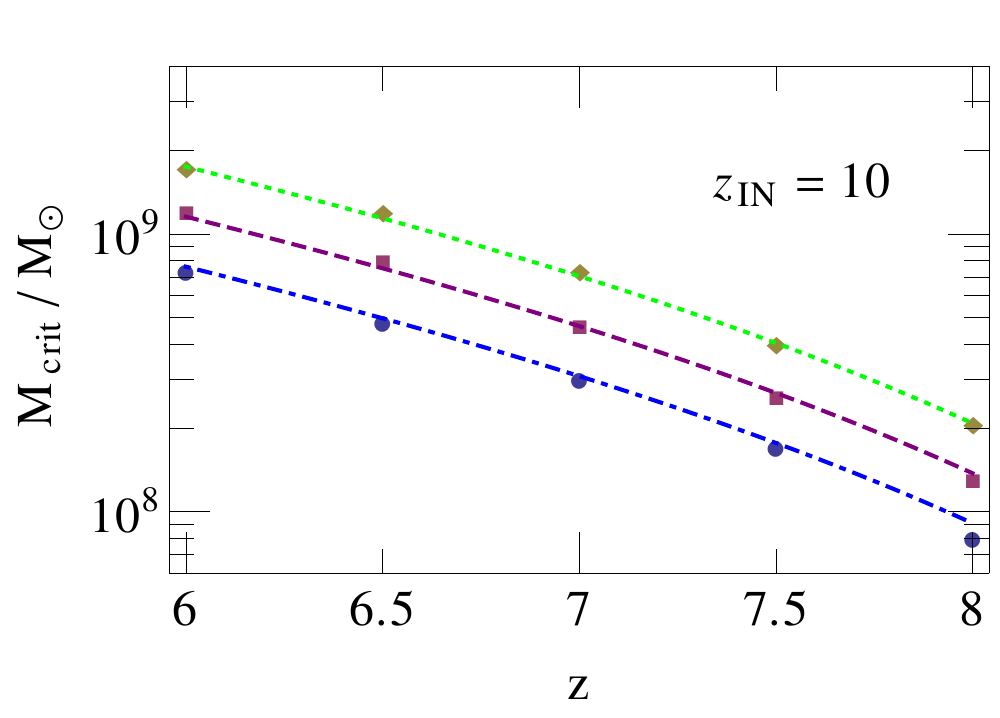}
\includegraphics[width=0.33\textwidth]{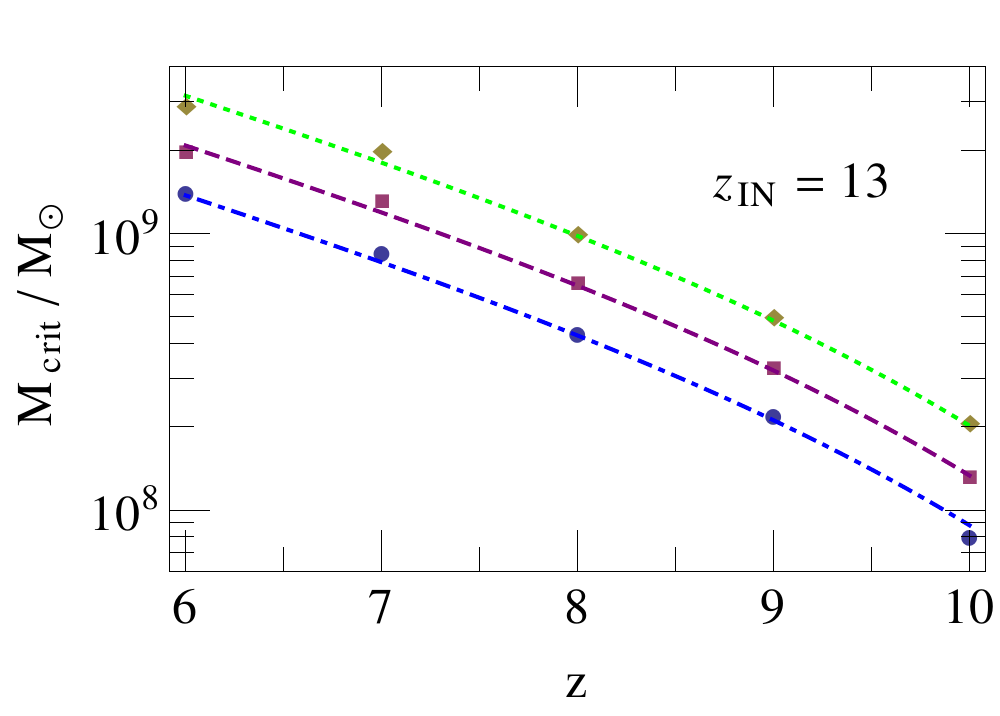}
\includegraphics[width=0.33\textwidth]{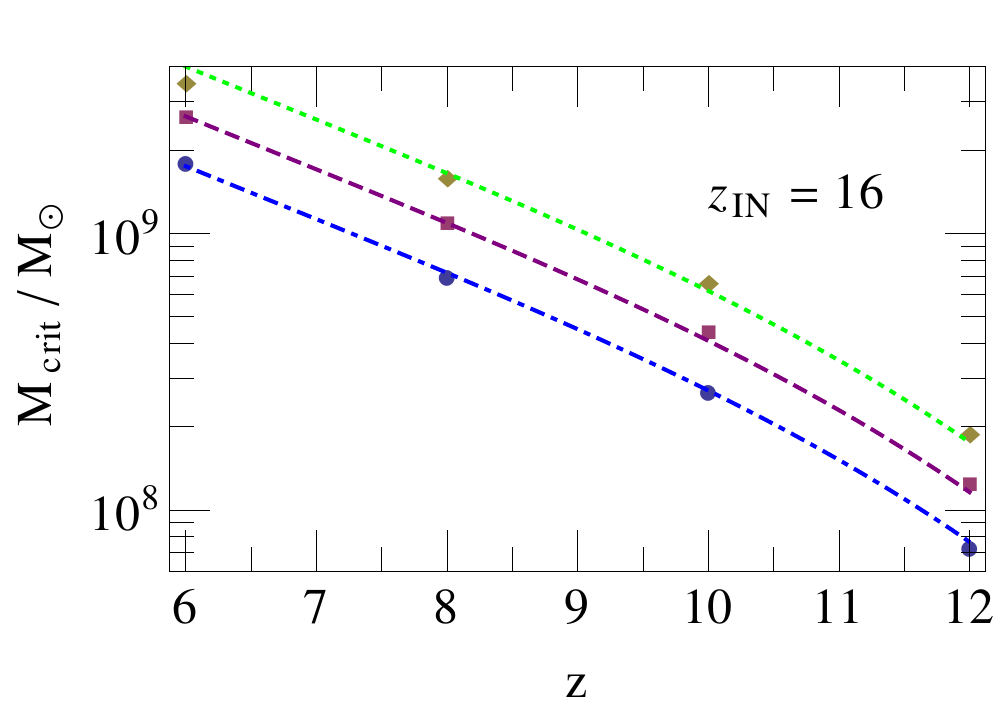}
}
\caption{Critical mass versus redshift with different choices of $z_{\rm IN}$; points show the output of the simulations and curves show the best fit of eq. \ref{eq:critical_mass_3} with three different values of $J_{\rm 21}$: \textit{Dot-dashed}: $J_{\rm 21}=0.01$. \textit{Dashed}: $J_{\rm 21}=0.1$. \textit{Dotted}: $J_{\rm 21}=1$.
\label{fig:crit_mass}
}
\vspace{-1\baselineskip}
\end{figure*}

Armed with our suite of collapse runs, we now present the two main results of this paper: a formula for the critical mass $M_{\rm crit}\left(J, z, z_{\rm IN}\right)$ (defined as the halo mass retaining half of its baryons; \S \ref{sec:crit_mass}), and its generalization to the halo baryon fraction $f_{\rm b}\left(M, J, z, z_{\rm IN} \right)$ (\S \ref{sec:f_b}).
Our approach allows us to {\it explicitly include dependences of these quantities on $J$ and $z_{\rm IN}$}.  This is important both for its generality, and also since all reasonable reionization models are very inhomogeneous (e.g. \citealt{MQS12, SM12b}).

\subsection{Critical Mass}
\label{sec:crit_mass}

As an ionization front sweeps through a patch of the IGM at $z=z_{\rm IN}$, the temperature of the gas is expected to jump from $T_{z>z_{\rm IN}} \sim 0$, to $T_{z<z_{\rm IN}} \sim10^4$ K (e.g. \citealt{BL01}), roughly within a sound-crossing time.
Using linear theory in such a scenario, \citet{GH98} (Appendix A) motivate the suppression of gas on scales smaller than the ``filtering'' wavenumber:
\begin{equation}
\label{eq:GH}
\frac{1}{k_F^2} = \frac{3}{10k_J^2} \left[  1 + 4\left( \frac{1+z}{1+z_{\rm IN}} \right)^{5/2} - 5 \left( \frac{1+z}{1+z_{\rm IN}} \right)^2  \right] ~ ,
\end{equation}
\noindent where $k_J$ corresponds to the classical (instantaneous) Jeans wavenumber at $T_{z<z_{\rm IN}}$.  Although there is no obvious motivation why this linear theory derivation should extend to collapsing and cooling structures, the filtering scale was shown to agree well with hydrodynamical simulations at high redshifts ($z\gsim6$; \citealt{Gnedin00, OGT08, NBM09}).

We note from eq. (\ref{eq:GH}), that the temperature dependence (encoded in the Jeans wavenumber) factors out.  Since the intensity of the UVB mainly affects the gas temperature, we consider a factorable expression $M_{\rm crit}(J,z,z_{\rm IN})=f(J)g(z,z_{\rm IN})$. This is confirmed by our simulations and also consistent with former ones (e.g. \citealt{MD08}), which motivate a power law form $f\left(J\right)\propto J_{\rm 21}^{\phantom{21}a}$.  Furthermore, we expect that the asymptotic limit of $M_{\rm crit}(z\ll z_{\rm IN})$ is approached independently of the precise value of $z_{\rm IN}$. Therefore we can write
$M_{\rm crit}\left(J, z, z_{\rm IN}\right)=J_{\rm 21}^{\phantom{21}a}g_{\rm 1}\left(z\right)g_{\rm 2}\left(z, z_{\rm IN}\right)$, 
with $g_{\rm 2}(z, z_{\rm IN}) \sim 1$ if $z\ll z_{\rm IN}$. Since the Jeans mass before reionization is much lower than the mass scales we are studying, we assume the additional limiting behaviour of $g_{\rm 2}(z, z_{\rm IN}) \rightarrow 0$ at $z \rightarrow z_{\rm IN}$.

Mindful of these trends, we assume the functional form:
\begin{equation}
\label{eq:critical_mass_3}
M_{\rm crit}=M_{\rm 0}J_{\rm 21}^{\phantom{21}a}\left(\frac{1+z}{10}\right)^{b}\left[1-\left(\frac{1+z}{1+z_{\rm IN}}\right)^{c}\right]^{d}
\end{equation}
where we treat $M_{\rm 0}$, $a$, $b$, $c$ and $d$ as fitting parameters.\footnote{The same form with parameters ($M_{\rm 0}$, $a$, $b$, $c$, $d$) = ($1.9\times 10^{9} M_{\odot}$, $0$, $-3/2$, $3/2$, $3$) is obtained under the assumption that the baryons get photo-evaporated over a sound crossing time at $10^4\text{ K}$ (c.f. \citealt{SIR04})
\begin{equation}
\nonumber
t_{\rm sc}=200\text{Myr}\left(\frac{M}{10^{8}M_{\odot}}\right)^{1/3}\left(\frac{1+z}{10}\right)^{-1}\left(\frac{\Omega_{\rm m}h^{2}}{0.15}\right)^{-1/3}
\end{equation}
and that the halo grows over a longer time-scale. In this scenario $M_{\rm crit}$ is obtained solving the equation $t_{\rm sc}\left(M_{\rm crit}\right)=t_{\rm H}\left(z\right)-t_{\rm H}\left(z_{\rm IN}\right)$, where $t_{\rm H}$ is the cosmic time.}  A $\chi^{2}$ minimization fitting to our simulation outputs results in:
\begin{equation}
\label{eq:fit_const}
\left(M_{\rm 0}, a, b, c, d\right)=\left(2.8\times 10^{9}M_{\odot}, 0.17, -2.1, 2.0, 2.5\right)
\end{equation}

Values of $M_{\rm crit}$ from our simulation runs are shown as points in Fig. \ref{fig:crit_mass}, together with the curves corresponding the analytic expression above (eq. \ref{eq:critical_mass_3}--\ref{eq:fit_const}). The relative errors of this fitting formula with respect to the simulation values are all $\lesssim 10\%$.
 We have tested the robustness of the fit by including additional data points with different choices of $z$ and $z_{\rm IN}$ (each spanning the range 7--19), and find that the best-fit parameters changed only within few percent of those in (\ref{eq:fit_const}).
Note that our results are not very sensitive to the intensity of the UVB: increasing $J_{\rm 21}$ by two orders of magnitude increases $M_{\rm crit}$ only by a factor $\sim 2$--3.  This weak dependence is expected given that the post-ionization collapse in this regime proceeds isothermally, and $T_{z<z_{\rm IN}}\sim10^4$K is only weakly dependent on $J$ \citep{DHRW03}.

The functional form for $M_{\rm crit}$ assumes a rapid transition at $z_{\rm IN}$, i.e. $T_{z>z_{\rm IN}} \ll T_{z<z_{\rm IN}}$.  However, earlier populations of X-ray sources could have pre-heated the IGM, making the change in the Jeans mass at $z_{\rm IN}$  less dramatic.  Of greater interest is the ability of the gas to cool efficiently, allowing it to continue contracting past the adiabatic stage, eventually forming stars.
It is thought that the first galaxies formed in halos with virial temperatures $T_{\rm vir} \ll 10^4$ K, governed by the molecular cooling channel.  However, these objects were very sensitive to highly uncertain feedback processes (e.g. \citealt{MBA03, KM05}), making it difficult to estimate the effective minimum halo mass required to host galaxies, $M_{\rm min}$, at epochs around and before $z_{\rm IN}$.  Star-forming galaxies need to be massive enough to both: (i) host gas in the presence of a UVB ($M>M_{\rm crit}$); and (ii) allow this gas to efficiently cool, collapse and form stars ($M>M_{\rm cool}$).  Therefore, a reasonable, general choice for $M_{\rm min}$ would be:
\begin{equation}
\label{eq:M_min}
M_{\rm min}=\max\left[M_{\rm cool}, M_{\rm crit}\right] ~ ,
\end{equation}
\noindent where $M_{\rm cool}$ corresponds to the effective cooling threshold (including feedback effects).

Our expression for $M_{\rm crit}$ is a good match to the ``filtering'' mass scale, as well as to numerical simulation at high-redshifts ($z\gsim6$; e.g. \citealt{OGT08}). However our expression has a steeper low-redshift evolution with respect to the simulations of \citet{OGT08}, over-predicting the value at $z=0$ by a factor of $\approx10$ (see their appendix B).  The particular realization of \citet{OGT08} prefers an $M_{\rm crit}$ which asymptotes to a constant virial temperature, i.e. $M_{\rm crit}\propto\left(1+z\right)^{-3/2}$.
A possible reason for this discrepancy is the fact that our simulations are missing 3D substructures, whose resulting potential wells would aid in the retention of gas.  Hence, we caution against over-interpreting eq. (\ref{eq:critical_mass_3}) at low-redshifts ($z\lsim5$), where halos increasingly grow by mergers and our simple, spherically-symmetric picture breaks down.  {\it In a companion paper \citep{SM12b}, we provide a general, analytic expression for the global average $\bar{M}_{\rm min}(z)$, considering a ``fiducial'' inhomogeneous reionization model and a constant $T_{\rm vir}$ low-redshift limit.}

\subsection{Halo Baryon Fraction}
\label{sec:f_b}

We now generalize the expression for the critical mass, obtaining the halo baryon fraction, $f_b$, normalized to the global mean, $\Omega_{\rm b}/\Omega_{\rm m}$.
By definition, $f_b(M=M_{\rm crit}) = 1/2$, with the asymptotic behaviour: $f_b(M \ll M_{\rm crit}) = 0$ and $f_b(M \gg M_{\rm crit}) = 1$.
We find our results are well-fit (with relative errors $\lesssim 5\%$), by the following simple expression \footnote{Our results are also well-fit with commonly-used expression suggested by \citet{Gnedin00} (his eq. 7, with $\alpha=1$).  However, as that functional form is also not physically motivated, we use the simpler expression in eq. (\ref{f_b}).}:
\begin{equation}
\label{f_b}
f_{\rm b}\left(M\right)=2^{-M_{\rm crit}/M} ~ .
\end{equation}
\noindent We note that this expression is a relatively steep function of $M$, thus justifying the common, step-function simplification of $f_b= 0\rightarrow1$ at $M_{\rm crit}$.

\begin{figure*}
\vspace{+0\baselineskip}
{
\includegraphics[width=0.33\textwidth]{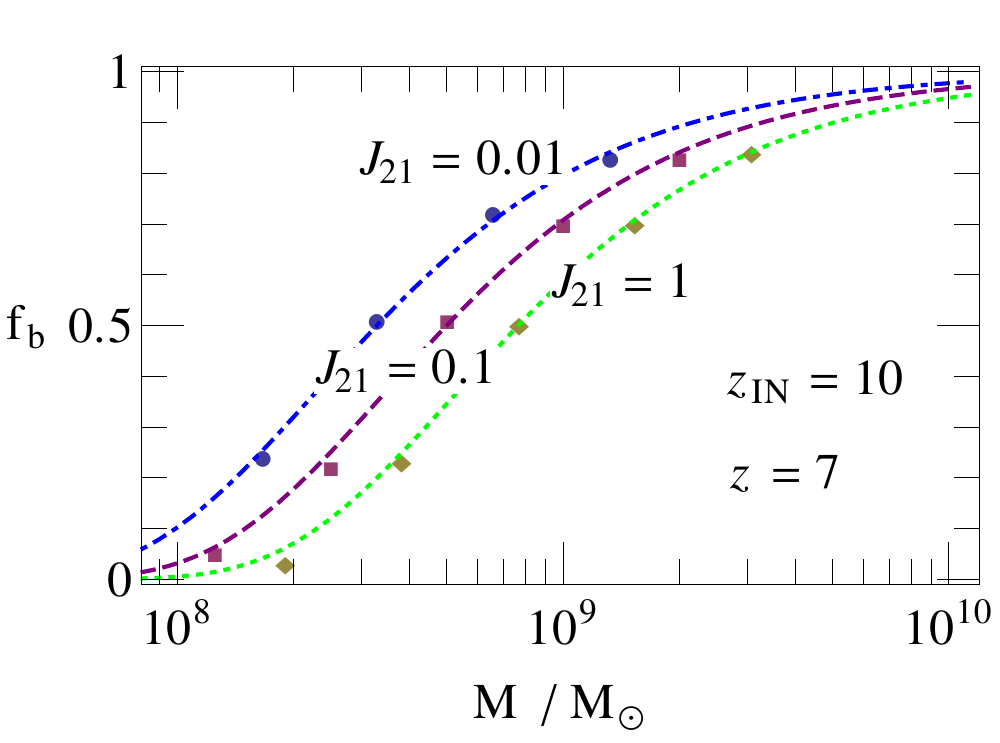}
\includegraphics[width=0.33\textwidth]{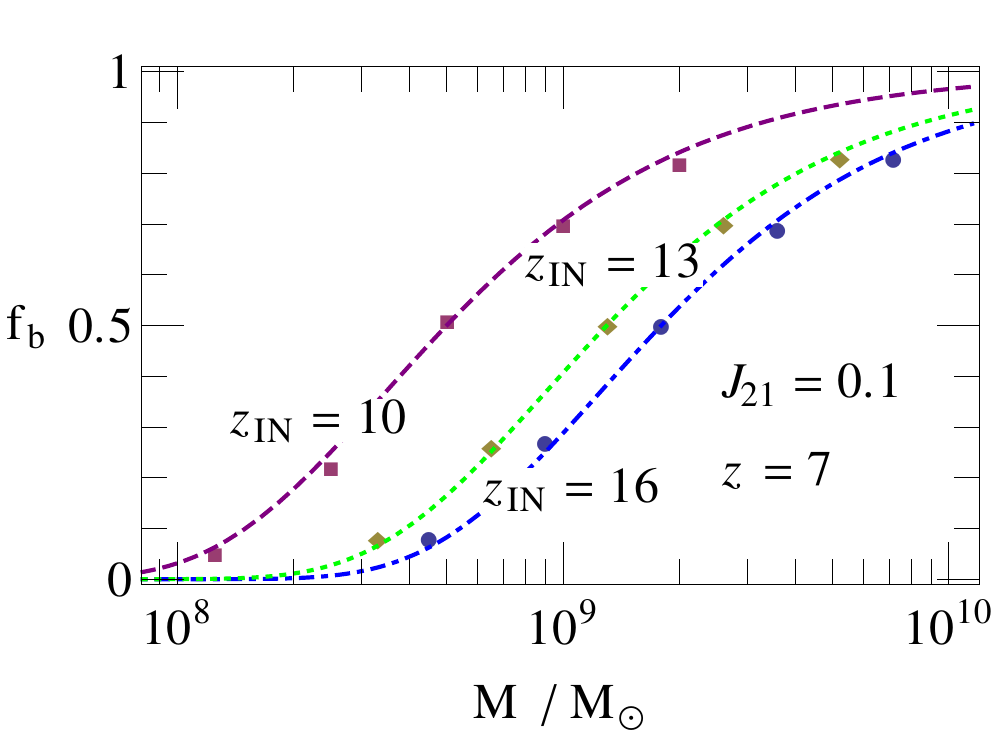}
\includegraphics[width=0.33\textwidth]{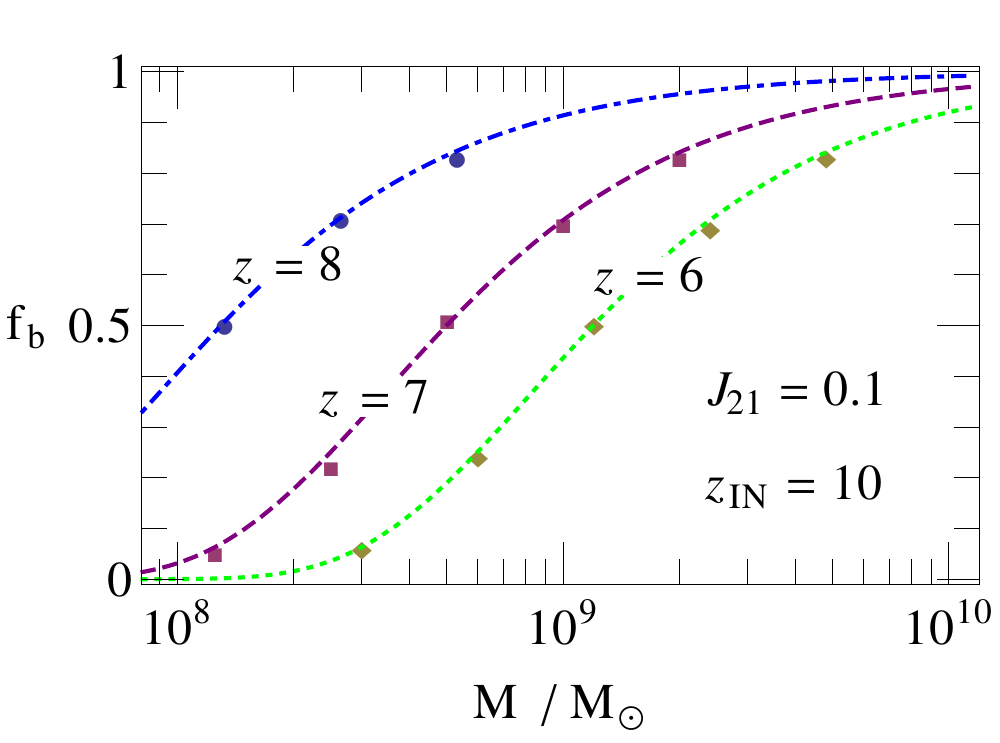}
}
\caption{The baryon fraction, $f_{\rm b}$, versus $M$ for halos with different reionization histories. Points show the output of the simulations and curves show our fitting formula in eq. (\ref{f_b}). We choose a ``fiducial'' parameter combination: $\left(z, z_{\rm IN}, J_{\rm 21}\right)=\left(7, 10, 0.1\right)$ and show variation with $J_{\rm 21}$, $z_{\rm IN}$, $z$ in the left, middle and right panels (respectively).
\label{fig:f_b}
}
\vspace{-1\baselineskip}
\end{figure*}

In Fig. \ref{fig:f_b} we show the baryon fraction of halos with different masses and reionization histories. Points show the output of the simulation and curves show our fitting formula. We present a ``fiducial'' model, $\left(z, z_{\rm IN}, J_{\rm 21}\right)=\left(7, 10, 0.1\right)$, varying one parameter in each panel.
In the left panel, we show the variation of $f_{\rm b}$ with $J_{\rm 21}$.  As we have already discussed this dependence is weak.
In the middle panel, we show the variation of $f_{\rm b}$ with $z_{\rm IN}$.  This dependence is stronger.  In particular, regions only recently exposed to a UVB do not exhibit significant baryonic suppression.
This is qualitatively supportive of earlier claims \citep{MD08} that UVB feedback does not significantly impact the progress of the advanced stages of reionization (note that this panel shows a ``maximal'' $z_{\rm IN}-z$ case of halos collapsing towards the very end of reionization, $z=7$).
These results also suggest that a potentially large scatter in the reionization redshift of halos (e.g. \citealt{ABAW09}) may have an important impact on the evolution of their gas.
Finally, in the right panel we show the variation of $f_{\rm b}$ with $z$.  The formation redshift has a very strong impact on the baryon fraction, consistent with claims that UVB feedback can be important in suppressing star formation post-reionization (e.g. \citealt{TW96, BLBC02}).


\subsection{Limitations of our Method}
\label{sec:disc}

Using spherically-symmetric collapse simulations allows us to rapidly explore a broad parameter space, explicitly quantifying the dependence of the halo's baryon content on its reionization history.  This is a critical improvement to previous work since reionization is likely to be very inhomogeneous.  Our general approach facilitates its inclusion in reionization models. Nevertheless, it suffers from several limitations:
\vspace{-0.3cm}
\begin{packed_enum}
\item {\it Evolving background:} Our parameter space only includes non-evolving UVB intensities.  As mentioned above, one might expect the ionizing background to decrease towards higher redshifts.  Observations of the Ly$\alpha$ forest post-reionization imply a fairly constant UVB over the range $2 \lsim z \lsim 5$ (e.g. \citealt{BH07}). Estimating its evolution towards higher redshifts is limited by our understanding of absorption systems and galaxy emissivity (e.g. \citealt{QOF11}).
  In any case, we show that feedback is fairly insensitive to the UVB intensity, so we do not expect any realistic evolution to affect our results.
\item {\it Self-shielding:} Our models do not include self-shielding, which could decrease the amount of ionizing radiation reaching the inner parts of the structures, thus reducing the impact of UVB feedback (e.g. \citealt{SU04}).
Spherically symmetric models likely overestimate this effect since ionizing photons can likely penetrate inside structures following low HI column density paths.  Furthermore, the UVB can strip away the outer layers of gas, rapidly exposing deeper layers which were previously self-shielded, with photo-evaporation occurring in roughly the same time-scale as expected in the optically-thin limit \citep{ISR05}.  Regardless, we find that the evolution of gas at the outskirts of halos plays a dominant role in governing the UVB feedback, consistent with previous claims (e.g. \citealt{OGT08}). 
\item {\it 3D structure:}  Although spherical symmetry is a decent approximation for early-collapsing structures (e.g. \citealt{BBKS86}),
this approximation becomes increasingly suspect towards lower redshifts.  Substructures form earlier than the collapse of the analogous spherical shell, making it easier for them to retain gas.
As discussed above, we suspect this to be an important consideration at $z\lsim5$ \citep{OGT08}.  Further progress is likely to involve suites of 3D simulations (including radiative transfer), guided by the functional forms motivated above.
\end{packed_enum}

\section{Conclusions}
\label{sec:concl}

Using a spherically symmetric simulation, we study the baryonic content of atomically-cooled galaxies exposed to a UVB.  Since the details of reionization are unclear, we explore a large parameter space of (i) halo mass; (ii) UVB intensity; (iii) redshift; (iv) redshift of UVB exposure.  The later is especially important given that reionization is likely very inhomogeneous.
We present an analytic expression for the characteristic or critical mass $M_{\rm crit}$, defined as the mass scale of halos retaining half their gas mass compared to the global mean.  We also generalize these results, obtaining a simple formula for the baryonic content of galaxies as a function of (i)--(iv) above.  Our results can be readily applied to models of inhomogeneous reionization and high-$z$ galaxy formation.

\vskip+0.3in

We thank Mark Dijkstra for providing and technical assistance with the code we have used as well as for interesting discussions. We thank Matthew McQuinn, Stuart Wyithe, Zoltan Haiman, and Andrea Ferrara for the helpful comments on early drafts of this paper.

\bibliographystyle{mn2e}
\bibliography{ms}

\end{document}